\documentclass[aps,twocolumn,showpacs,preprintnumbers,floats]{revtex4}\def\@cite#1#2{\textsuperscript{[{#1\if@tempswa , #2\fi}]}}

\usepackage{graphicx}
\usepackage{amsmath}
\usepackage{amsfonts}
\usepackage{amssymb}
\usepackage{color}
\usepackage{subfigure}
\usepackage{epsfig}
\usepackage{morefloats}
\usepackage{multirow}
\usepackage{graphicx,booktabs}
\usepackage{mathrsfs}
\usepackage{txfonts}
\usepackage{indentfirst}
\usepackage{graphicx,booktabs}

\usepackage{longtable,lscape}

\newcommand{\vsig}{\mbox{\boldmath$\sigma$\unboldmath}}

\usepackage{comment}

\begin{document}


\title{Possible interpretation of the newly observed $\Omega_b$ states}

\author{Li-Ye Xiao$^{1}$~\footnote {E-mail: lyxiao@ustb.edu.cn}, Kai-Lei Wang$^{2}$~\footnote {E-mail: wangkaileicz@foxmail.com}, Ming-Sheng Liu$^{3,4}$~\footnote {E-mail: liumingsheng0001@126.com} and Xian-Hui Zhong$^{3,4}$~\footnote {E-mail: zhongxh@hunnu.edu.cn}}

\affiliation{ 1)School of Mathematics and Physics, University of Science and Technology Beijing,
Beijing 100083, China}
\affiliation{ 2) Department
of Electronic Information and Physics, Changzhi University, Changzhi, Shanxi,046011,China}  
\affiliation{ 3) Department of
Physics, Hunan Normal University, and Key Laboratory of
Low-Dimensional Quantum Structures and Quantum Control of Ministry
of Education, Changsha 410081, China }
\affiliation{ 4) Synergetic
Innovation Center for Quantum Effects and Applications (SICQEA),
Hunan Normal University, Changsha 410081, China}


\begin{abstract}

Inspired by the newly observed $\Omega_b$ states by the LHCb Collaboration, we investigate the two-body
strong decays of the low-lying $\lambda$-mode $\Omega_b$ baryons up to $N=2$ shell using the chiral quark model within the $j$-$j$ coupling scheme.
Our results indicate that: (i) the newly observed states $\Omega_b(6316)^-$ and $\Omega_b(6330)^-$ are good candidates of the light spin $j=1$ states, while the spin-parity $J^P=1/2^-$ and $J^P=3/2^-$ cannot be distinguished.
The other two states, $\Omega_b(6340)$ and $\Omega_b(6350)$ mostly correspond to the light spin $j=2$ states, while the spin-parity $J^P=3/2^-$ and $J^P=5/2^-$ cannot be distinguished as well. (ii) The $2S$ states with spin-parity $J^P=1/2^+$ and $J^P=3/2^+$, respectively, might be narrow states with a width of $\Gamma<2$ MeV. (iii) The $1D$ states are not broad and the total decay widths vary from several to dozens of MeV, which have a good potential to be observed in future experiments.

\end{abstract}


\maketitle

\section{Introduction}

Since the heavy quark provides a ``flavour tag" that may give insight into the systematics of hadron resonances and the mechanism of confinement, the studies of hadrons containing heavy quarks ($c$ or $b$) are important topics in hadron physics. It is evident that heavy-quark hadrons play an important role in our understanding of QCD. Since the first observation of the beauty baryon $\Lambda_b$ at the CERN $p\bar{p}$ collider~\cite{Albajar:1991sq}, many singly heavy baryons were observed by various world wide experimental facilities in the past three decades~\cite{Tanabashi:2018oca}. In recent years, there have been specially fruitful in the observation of the heavy baryon states~\cite{Aaij:2017vbw,Aaij:2017nav,Yelton:2017qxg,Aaij:2018tnn,Aaij:2018yqz,Aaij:2019amv,Aaij:2017ueg}. The rapidly growth of the experimental progress attracts a
a great deal of attention from the hadron physics community.

Very recently, in the $\Xi^0_bK^-$ mass spectrum obtained using $pp$ collisions the LHCb Collaboration reported four narrow peaks~\cite{Aaij:2020cex},
\begin{eqnarray}
m[\Omega_b(6316)^-]=6315.64\pm0.31\pm0.07\pm0.50~\text{MeV},\nonumber\\
\Gamma[\Omega_b(6316)^-]<2.8(4.2)~~\text{MeV},\\
m[\Omega_b(6330)^-]=6330.30\pm0.28\pm0.07\pm0.50~\text{MeV},\nonumber\\
\Gamma[\Omega_b(6330)^-]<3.1(4.7)~~\text{MeV},\\
m[\Omega_b(6340)^-]=6339.71\pm0.26\pm0.05\pm0.50~\text{MeV},\nonumber\\
\Gamma[\Omega_b(6340)^-]<1.5(1.8)~~\text{MeV},\\
m[\Omega_b(6350)^-]=6349.88\pm0.35\pm0.05\pm0.50~\text{MeV},\nonumber\\
\Gamma[\Omega_b(6316)^-]<2.8(3.2)~~\text{MeV},
\end{eqnarray}
where the uncertainties are statistical, systematic and the limited knowledge of the $\Xi^0_b$ mass, respectively. The natural widths $\Gamma$ correspond to $90\%$ ($95\%$) confidence level upper limits.

Before the LHCb's measurement~\cite{Aaij:2020cex}, there exist many theoretical calculations of the mass spectrum of the excited $\Omega_b$ states with various quark models~\cite{Yoshida:2015tia,Thakkar:2016dna,Ebert:2011kk,Santopinto:2018ljf,Yang:2017qan,Roberts:2007ni,Garcilazo:2007eh,Ebert:2007nw}, Regge phenomenology~\cite{Wei:2016jyk}, QCD two-point sum rule method~\cite{Agaev:2017jyt,Mao:2015gya}, etc. We collect the theoretical predictions in Table~\ref{mass-spectrum}. From the table, the four peaks observed by the LHCb Collaboration~\cite{Aaij:2020cex} are in the predicted mass region of the $1P$ wave excitations~\cite{Yoshida:2015tia,Thakkar:2016dna,Ebert:2011kk,Santopinto:2018ljf,Roberts:2007ni,Garcilazo:2007eh,Ebert:2007nw}. In addition, the possibility as $2S$ excitations cannot be excluded absolutely based simply on the predicted masses~\cite{Yang:2017qan,Garcilazo:2007eh}. There are also some discussions of the strong decay properties of the excited $\Omega^-_b$ states. In the framework of the QCD sum rule method~\cite{Agaev:2017ywp}, the $1P$- and $2S$-wave states with spin-1/2 and spin-3/2 are predicted to be narrow resonances. Within an harmonic oscillator based model, the authors in Ref.~\cite{Santopinto:2018ljf} obtained that the $1P$-wave states should be narrow resonances as well.
With a chiral quark model our group predicted that in the $L$-$S$ coupling scheme the $1P$-wave states with spin-parity $J^P=3/2^-$ and $J^P=5/2^-$ should be narrow states with a width of a few MeV, while two $J^P=1/2^-$ $1P$-wave states $| ^2P_{1/2}\rangle$ and $| ^4P_{1/2}\rangle$ are most likely to be broad states with a width of $\Gamma\sim (50-100)$ MeV ~\cite{Wang:2017kfr}; however, in the $j$-$j$ coupling scheme only the $j=0$ $P$-wave state with $J^P=1/2^-$ is a broad state, the other four $1P$-wave states should be narrow states~\cite{Wang:2018fjm}. Recently, Liang et al.~\cite{Liang:2020hbo} assigned the observed four peaks as the $1P$ wave states within the quark pair creation model. The similar results were obtained in Refs.~\cite{Wang:2020pri,Chen:2020mpy}. In addition, the possible interpretations of the four peaks as the meson-baryon molecular states were presented as well in Ref.~\cite{Liang:2020dxr}.

\begin{table*}[ht]
\caption{The predicted masses (MeV) of the $\Omega^-_b$ states within various phenomenal models and methods. $|J^P,j\rangle~(nl)$ denotes the state within the $j$-$j$ coupling scheme.} \label{mass-spectrum}
\begin{tabular}{cccccccccccccccc }\hline\hline
$|^{2S+1}L_J\rangle(NL)$ ~~~&NRQM~\cite{Yoshida:2015tia} ~~&hCQM~\cite{Thakkar:2016dna}~~&RQM~\cite{Ebert:2011kk}~~&QM~\cite{Santopinto:2018ljf}
~~&ChQM~\cite{Yang:2017qan}~~&QM~\cite{Roberts:2007ni}~~&CQM~\cite{Garcilazo:2007eh}~~&CQM~\cite{Ebert:2007nw} ~~&Exp~\cite{Tanabashi:2018oca} \\
\hline
$|J^P=\frac{1}{2}^+,1\rangle(1S)$          ~~&6076 ~~&6048~~&6064~~&$6061\pm15$~~&6045~~&6081~~&6037~~&6065 ~~&6046 \\
$|J^P=\frac{3}{2}^+,1\rangle(1S)$          ~~&6094 ~~&6086~~&6088~~&$6082\pm20$~~&6045~~&6102~~&6090~~&6088 ~~&\\
$|J^P=\frac{1}{2}^-,0\rangle(1P)$~~&6333 ~~&6338~~&6339~~&$6305\pm15$~~&6288~~&6301~~&6278~~&6352 ~~& \\
$|J^P=\frac{1}{2}^-,1\rangle(1P)$~~&6340 ~~&6343~~&6330~~&$6317\pm19$~~&6288~~&6312~~&6373~~&6361 ~~& \\
$|J^P=\frac{3}{2}^-,1\rangle(1P)$~~&6336 ~~&6328~~&6340~~&$6313\pm15$~~&6288~~&6304~~&6278~~&6330 ~~& \\
$|J^P=\frac{3}{2}^-,2\rangle(1P)$~~&6344 ~~&6333~~&6331~~&$6325\pm19$~~&6288~~&6311~~&6373~~&6351 ~~& \\
$|J^P=\frac{5}{2}^-,2\rangle(1P)$~~&6345 ~~&6320~~&6334~~&$6338\pm20$~~&6288~~&6311~~&$\cdot\cdot\cdot$ ~~&6336 ~~& \\
$|J^P=\frac{1}{2}^+,1\rangle(2S)$          ~~&6517 ~~&6455~~&6450~~&$\cdot\cdot\cdot$~~&6384~~&6472~~&6367~~&6440 ~~&\\
$|J^P=\frac{3}{2}^+,1\rangle(2S)$          ~~&6528 ~~&6481~~&6461~~&$\cdot\cdot\cdot$~~&6384~~&6478~~&6398~~&6450 ~~& \\
$|J^P=\frac{1}{2}^+,1\rangle(1D)$~~&6561 ~~&6601~~&6540~~&$\cdot\cdot\cdot$~~&$\cdot\cdot\cdot$~~&$\cdot\cdot\cdot$~~&$\cdot\cdot\cdot$~~&6526 ~~& \\
$|J^P=\frac{3}{2}^+,1\rangle(1D)$~~&6559 ~~&6583~~&6549~~&$\cdot\cdot\cdot$~~&$\cdot\cdot\cdot$~~&$\cdot\cdot\cdot$~~&$\cdot\cdot\cdot$~~&6518 ~~& \\
$|J^P=\frac{3}{2}^+,2\rangle(1D)$~~&6559 ~~&6589~~&6530~~&$\cdot\cdot\cdot$~~&$\cdot\cdot\cdot$~~&$\cdot\cdot\cdot$~~&$\cdot\cdot\cdot$~~&6520 ~~& \\
$|J^P=\frac{5}{2}^+,2\rangle(1D)$~~&6561 ~~&6567~~&6529~~&$\cdot\cdot\cdot$~~&$\cdot\cdot\cdot$~~&6492~~&$\cdot\cdot\cdot$~~&6512 ~~& \\
$|J^P=\frac{5}{2}^+,3\rangle(1D)$~~&6566 ~~&6573~~&6520~~&$\cdot\cdot\cdot$~~&$\cdot\cdot\cdot$~~&6494~~&$\cdot\cdot\cdot$    ~~&6490 ~~& \\
$|J^P=\frac{7}{2}^+,3\rangle(1D)$~~&$\cdot\cdot\cdot$ ~~&6553~~&6517~~&$\cdot\cdot\cdot$~~&$\cdot\cdot\cdot$~~&6497~~&$\cdot\cdot\cdot$    ~~&6502 ~~& \\
\hline\hline
\end{tabular}
\end{table*}

We notice that for the singly heavy flavour quark systems, proper consideration of the heavy quark symmetry
is necessary, thus, the states may more favor the $j$-$j$ coupling scheme. With this coupling scheme, we have studied the strong decays of the $1P$-wave singly bottom heavy baryons in our previous work~\cite{Wang:2018fjm}. In the present work, we further conduct a systematical study of the strong decay properties of the low-lying $\lambda$-mode $\Omega_b$ states in the $j$-$j$ coupling scheme, and attempt to decode the inner structures of the four peaks observed by the LHCb Collaboration~\cite{Aaij:2020cex}. Our theoretical results suggest that both the $\Omega_b(6316)^-$ and $\Omega_b(6330)^-$ are good candidates of the light spin $j=1$ states, while the spin-parity $J^P=1/2^-$ and $J^P=3/2^-$ cannot be distinguished due to the limited data. The other two states, $\Omega_b(6340)^-$ and $\Omega_b(6350)^-$ are good candidates of the light spin $j=2$ states, while the spin-parity $J^P=3/2^-$ and $J^P=5/2^-$ cannot be distinguished as well. In addition, we notice that the light spin $j=0$ states with the spin-parity $J^P=1/2^-$ are most likely to be a moderate state with a width of $\Gamma\sim100$ MeV, and the $\Xi_b K$ decay channel almost saturates its total decay widths. For the $2S$-wave $\Omega_b$ states with $J^P=1/2^+$ and $J^P=3/2^+$, they might be narrow states with a width of $\Gamma<2$ MeV. We notice that the $1D$ states are not broad and the total decay widths vary from several to dozens of MeV, which have a good potential to be observed in future experiments.

This paper is organized as follows. In Sec. II we give a brief
introduction of the strong decay model and the relationship of the singly-heavy baryon spectrum between the $j$-$j$ coupling scheme and the $L$-$S$ coupling scheme. We discuss the strong decays of the low-lying $1P$-, $2S$-, and $1D$-wave $\Omega_b$ states within the $j$-$j$ coupling scheme in Sec. III and summarize our results in
Sec. IV.

\section{ the model}

In this work we study the strong decay properties with the chiral quark model. This model has been successfully
applied to study the strong decays of baryons  and heavy-light mesons~\cite{Zhong:2007gp,Liu:2012sj,Wang:2019uaj,Yao:2018jmc,Wang:2017kfr,Wang:2018fjm,Wang:2017hej,Xiao:2017udy,Xiao:2013xi,Xiao:2018pwe,
Liu:2019wdr,Zhong:2008kd,Zhong:2009sk,Zhong:2010vq,Xiao:2014ura}. In the chiral quark model, the effective low energy quark-pseudoscalar-meson coupling in the SU(3) flavor basis at tree level is described by~\cite{Manohar:1983md}
\begin{eqnarray}\label{coup}
H_m=\sum_j
\frac{1}{f_m}\bar{\psi}_j\gamma^{j}_{\mu}\gamma^{j}_{5}\psi_j\vec{\tau}\cdot
\partial^{\mu}\vec{\phi}_m,
\end{eqnarray}
where $\psi_j$ corresponds to the $j$th quark field in a baryon. $f_m$ stands for the pseudoscalar meson decay constant. $\phi_m$ denotes the pseudoscalar meson octet.

To match the nonrelativistic harmonic oscillator spatial wave function $^N\Psi_{LL_z}$ in the calculations, we adopt a nonrelativistic form of the quark-pseudoscalar coupling and obtain~\cite{Zhao:2002id,Li:1994cy,Li:1997gd}
\begin{eqnarray}\label{non-relativistic-expans}
H^{nr}_{m}&=&\sum_j\Big\{\frac{\omega_m}{E_f+M_f}\vsig_j\cdot
\textbf{P}_f+ \frac{\omega_m}{E_i+M_i}\vsig_j \cdot
\textbf{P}_i \nonumber\\
&&-\vsig_j \cdot \textbf{q} +\frac{\omega_m}{2\mu_q}\vsig_j\cdot
\textbf{p}'_j\Big\}I_j e^{-i\mathbf{q}\cdot \mathbf{r}_j},
\end{eqnarray}
where $(E_i,~M_i,~\mathbf{p}_i)$ and $(E_f,~M_f,~\mathbf{p}_f)$ represent the energy, mass and three-vector momentum of the initial and final baryon, respectively. ($\omega_m,~\mathbf{q}$) are the energy and three-vector momentum of the final light pseudoscalar meson. $\mathbf{p}'_j=\mathbf{p}_j-(m_j/M)\mathbf{P}_{\text{c.m.}}$ is the internal momentum of the $j$th quark in the baryon rest frame. $\mu_q$ stands for a reduced mass expressed as $1/\mu_q=1/m_j+1/m'_j$, and $\mathbf{\sigma}_j$ denotes the Pauli spin vector on the $j$th quark. The isospin operator $I_j$ associated with $K$ meson is given by
\begin{equation}
I_{j}=\begin{cases}
                             a^{\dagger}_j(u)a_j(s)     &$for$~K^{-},\\
        a^{\dagger}_j(d)a_j(s)   &$for$~\bar{K}^0.
       \end{cases}
\end{equation}
Here, $a_j^{\dagger}(u,d)$ and $a_j(s)$ are the creation and annihilation operator for the $u,~d$ and $s$ quarks on $j$th quark.

Then the partial decay width for the emission of a light pseudoscalar meson is
\begin{equation}\label{dww}
\Gamma=\left(\frac{\delta}{f_m}\right)^2\frac{(E_f +M_f)|\mathbf{q}|}{4\pi
M_i}\frac{1}{2J_i+1}\sum_{J_{iz}J_{fz}}|\mathcal{M}_{J_{iz},J_{fz}}|^2,
\end{equation}
where $J_{iz}$ and $J_{fz}$ are the third components of the total angular momenta of the initial and final baryons, respectively. $\delta$ denotes a global parameter accounting for the strength of the quark-meson couplings. $\mathcal{M}_{J_{iz},J_{fz}}$ stands for the transition amplitude.

According to our previous works~\cite{Wang:2018fjm,Wang:2017hej} together with the nature of the newly observed four peaks~\cite{Aaij:2020cex}, the physical states of the singly heavy baryons seem more favor the $j$-$j$ coupling scheme. Thus, in the present work we study the strong decay properties within this coupling scheme.

In the $L$-$S$ coupling scheme, the states are
constructed by~\cite{Roberts:2007ni}
\begin{equation}
\left|^{2S+1}L_{J}\right\rangle = \left|\left[\left(l_{\rho} l_{\lambda}\right)_L\left(s_{\rho}s_Q\right)_S\right]_{J^P}\right\rangle,
\end{equation}
where $l_{\rho}$ and $l_{\lambda}$ are the quantum numbers of
the orbital angular momenta $\mathbf{l}_{\rho}$ and $\mathbf{l}_{\lambda}$ for the $\rho$- and $\lambda$-mode oscillators, respectively.
$L$ corresponds to the quantum number of the total orbital angular momentum $\mathbf{L}=\mathbf{l}_{\rho}+\mathbf{l}_{\lambda}$, which determine the parity of a state by $P=(-1)^{l_{\rho}+l_{\lambda}}$. $s_{\rho}$ and $s_Q$ stand for the quantum numbers of the total spin of the two light quarks $\mathbf{s}_{\rho}$ and the spin of the heavy quark $\mathbf{s}_Q$, respectively.
$S$ is the quantum number of the total spin angular momentum $\mathbf{S}=\mathbf{s}_{\rho}+\mathbf{s}_Q$. According to the quark model classification, there are five $\lambda$-mode $1P$-wave states, six $\lambda$-mode $1D$-wave states and two $2S$-wave states, and their corresponding quantum numbers have been collected in Table~\ref{LS}.

\begin{table}[htp]
\begin{center}
\caption{\label{LS}  The classifications of the low-lying $\lambda$-mode $1P$-, $1D$- and $2S$-wave states within the $L$-$S$ coupling scheme.}
\begin{tabular}{p{2.5cm}p{0.5cm}p{0.5cm}p{0.5cm}p{0.5cm}p{0.5cm}p{0.5cm}p{0.5cm}p{0.5cm}p{0.5cm}p{0.5cm}p{0.5cm} |}\hline\hline
$|^{2S+1}L_J \rangle$      &$J^P$                       &$n_{\rho}$  &$l_{\rho}$ &$n_{\lambda}$  &$l_{\lambda}$  &$L$    &$s_{\rho}$   &$s_{Q}$    &$S$     \\
\hline
$|^2P_{\lambda}~\frac{1}{2}^-\rangle(1P)$       &$\frac{1}{2}^-$  & 0    & 0    & 0      & 1          &1          &1       &$\frac{1}{2}$  &$\frac{1}{2}$         \\
$|^2P_{\lambda}~\frac{3}{2}^-\rangle(1P)$       &$\frac{3}{2}^-$  & 0    & 0    & 0      & 1          &1          &1       &$\frac{1}{2}$  &$\frac{1}{2}$              \\
$|^4P_{\lambda}~\frac{1}{2}^-\rangle(1P)$       &$\frac{1}{2}^-$  & 0    & 0    & 0      & 1          &1          &1       &$\frac{1}{2}$  &$\frac{3}{2}$           \\
$|^4P_{\lambda}~\frac{3}{2}^-\rangle(1P)$       &$\frac{3}{2}^-$  & 0    & 0    & 0      & 1          &1          &1       &$\frac{1}{2}$  &$\frac{3}{2}$            \\
$|^4P_{\lambda}~\frac{5}{2}^-\rangle(1P)$       &$\frac{5}{2}^-$  & 2    & 0    & 0      & 0          &1          &1       &$\frac{1}{2}$  &$\frac{3}{2}$            \\
$|^2S~\frac{1}{2}^+\rangle(2S)$                 &$\frac{1}{2}^+$  & 0    & 0    & 1      & 0          &0          &1       &$\frac{1}{2}$  &$\frac{1}{2}$            \\
$|^4S~\frac{3}{2}^+\rangle(2S)$                 &$\frac{3}{2}^+$  & 0    & 0    & 1      & 0          &0          &1       &$\frac{1}{2}$  &$\frac{3}{2}$            \\
$|^2D_{\lambda\lambda}~\frac{3}{2}^+\rangle(1D)$&$\frac{3}{2}^+$  & 0    & 0    & 0      & 2          &2          &1       &$\frac{1}{2}$  &$\frac{1}{2}$         \\
$|^2D_{\lambda\lambda}~\frac{5}{2}^+\rangle(1D)$&$\frac{5}{2}^+$  & 0    & 0    & 0      & 2          &2          &1       &$\frac{1}{2}$  &$\frac{1}{2}$         \\
$|^4D_{\lambda\lambda}~\frac{1}{2}^+\rangle(1D)$&$\frac{1}{2}^+$  & 0    & 0    & 0      & 2          &2          &1       &$\frac{1}{2}$  &$\frac{3}{2}$         \\
$|^4D_{\lambda\lambda}~\frac{3}{2}^+\rangle(1D)$&$\frac{3}{2}^+$  & 0    & 0    & 0      & 2          &2          &1       &$\frac{1}{2}$  &$\frac{3}{2}$         \\
$|^4D_{\lambda\lambda}~\frac{5}{2}^+\rangle(1D)$&$\frac{5}{2}^+$  & 0    & 0    & 0      & 2          &2          &1       &$\frac{1}{2}$  &$\frac{3}{2}$         \\
$|^4D_{\lambda\lambda}~\frac{7}{2}^+\rangle(1D)$&$\frac{7}{2}^+$  & 0    & 0    & 0      & 2          &2          &1       &$\frac{1}{2}$  &$\frac{3}{2}$         \\
\hline\hline
\end{tabular}
\end{center}
\end{table}

\begin{table}[htp]
\begin{center}
\caption{\label{JJ}  The classifications of the low-lying $\lambda$-mode $1P$-, $1D$- and $2S$-wave states within the $j$-$j$ coupling scheme.}
\begin{tabular}{p{2.5cm}p{0.5cm}p{0.5cm}p{0.5cm}p{0.5cm}p{0.5cm}p{0.5cm}p{0.5cm}p{0.5cm}p{0.5cm}p{0.5cm}p{0.5cm} |}\hline\hline
$|J^P,j \rangle$      &$J^P$            &$j$  &$n_{\rho}$ &$\ell_{\rho}$  &$n_{\lambda}$  &$\ell_{\lambda}$    &$L$   &$s_{\rho}$    &$s_{Q}$                     \\ \hline
$|J^P=\frac{1}{2}^-,0\rangle$ ($1P$)   &$\frac{1}{2}^-$  & 0    & 0    & 0      & 0          &1          &1       &1         &$\frac{1}{2}$         \\
$|J^P=\frac{1}{2}^-,1\rangle$ ($1P$)   &$\frac{1}{2}^-$  & 1    & 0    & 0      & 0          &1          &1       &1         &$\frac{1}{2}$              \\
$|J^P=\frac{3}{2}^-,1\rangle$ ($1P$)   &$\frac{3}{2}^-$  & 1    & 0    & 0      & 0          &1          &1       &1         &$\frac{1}{2}$           \\
$|J^P=\frac{3}{2}^-,2\rangle$ ($1P$)   &$\frac{3}{2}^-$  & 2    & 0    & 0      & 0          &1          &1       &1         &$\frac{1}{2}$            \\
$|J^P=\frac{5}{2}^-,2\rangle$ ($1P$)   &$\frac{5}{2}^-$  & 2    & 0    & 0      & 0          &1          &1       &1         &$\frac{1}{2}$            \\
$|J^P=\frac{1}{2}^+,1\rangle$ ($2S$)  &$\frac{1}{2}^+$  & 1    & 0    & 0      & 1          &0          &0       &1         &$\frac{1}{2}$            \\
$|J^P=\frac{3}{2}^+,1\rangle$ ($2S$)  &$\frac{3}{2}^+$  & 1    & 0    & 0      & 1          &0          &0       &1         &$\frac{1}{2}$            \\
$|J^P=\frac{1}{2}^+,1\rangle$ ($1D$)  &$\frac{1}{2}^+$  & 1    & 0    & 0      & 0          &2          &2       &1         &$\frac{1}{2}$            \\
$|J^P=\frac{3}{2}^+,1\rangle$ ($1D$)  &$\frac{3}{2}^+$  & 1    & 0    & 0      & 0          &2          &2       &1         &$\frac{1}{2}$            \\
$|J^P=\frac{3}{2}^+,2\rangle$ ($1D$)  &$\frac{3}{2}^+$  & 2    & 0    & 0      & 0          &2          &2       &1         &$\frac{1}{2}$            \\
$|J^P=\frac{5}{2}^+,2\rangle$ ($1D$)  &$\frac{5}{2}^+$  & 2    & 0    & 0      & 0          &2          &2       &1         &$\frac{1}{2}$            \\
$|J^P=\frac{5}{2}^+,3\rangle$ ($1D$)  &$\frac{5}{2}^+$  & 3    & 0    & 0      & 0          &2          &2       &1         &$\frac{1}{2}$            \\
$|J^P=\frac{7}{2}^+,3\rangle$ ($1D$)  &$\frac{7}{2}^+$  & 3    & 0    & 0      & 0          &2          &2       &1         &$\frac{1}{2}$            \\
\hline\hline
\end{tabular}
\end{center}
\end{table}

Under the heavy-quark symmetry limit~\cite{Cheng:2015iom}, the states within the $j$-$j$ coupling scheme are
constructed by
\begin{eqnarray}
\left|J^P,j\right\rangle = \left|\left\{\left[\left(\ell_\rho \ell_\lambda\right)_L s_{\rho}\right]_js_Q\right\}_{J^P}\right\rangle.
\end{eqnarray}
Here, the total orbital angular momentum $\mathbf{L}=\mathbf{l}_{\rho}+\mathbf{l}_{\lambda}$ couples to the total spin of the two light quarks $\mathbf{s}_{\rho}$, and then gets $\mathbf{j}=\mathbf{L}+\mathbf{s}_{\rho}$. $\mathbf{j}$ couples to the spin of the heavy quark $\mathbf{s}_{Q}$, and gives the total angular momentum $\mathbf{J}=\mathbf{j}+\mathbf{s}_Q$. Similarly, with this coupling scheme the low-lying $\lambda$-mode $1P$-, $1D$- and $2S$-wave states are listed in Table~\ref{JJ}.

The states within the $j$-$j$ coupling scheme can be expressed as linear combinations of the states within the $L$-$S$ coupling scheme by~\cite{Roberts:2007ni}
\begin{eqnarray}
|\{[(l_\rho l_\lambda)_Ls_{\rho}]_js_Q\}_J\rangle =(-1)^{L+s_\rho+1/2+J}\sqrt{2j+1} \sum_{S}\sqrt{2S+1}  ~~\nonumber\\
\begin{Bmatrix}L &s_\rho &j\\s_Q &J &S \end{Bmatrix} |[(l_\rho l_\lambda)_L(s_{\rho}s_Q)_S]_J\rangle  . ~~~~~~~~~~~\label{Rela}
\end{eqnarray}

This relationship between the two different coupling schemes indicates that the states may well be mixed with the same spin-parity $J^P$ within the $L$-$S$ coupling scheme for the singly heavy baryons. The mixing angles can be estimated by Eq.~(\ref{Rela}).

In this work, the standard quark model parameters are adopted. Namely, we set $m_u=m_d=330$ MeV, $m_s=450$ MeV, and $m_b=5000$ MeV for the constituent quark masses. The decay constants for $K$ is taken as $f_{K}=160$ MeV. The harmonic oscillator space-wave functions are adopted to describe the spatial wave function of the initial and final baryons, and the harmonic oscillator parameter $\alpha_{\rho}$ in the wave functions for $ds/us$ and $ss$ systems are taken as $\alpha_{\rho}=420$ and 440 MeV, respectively. Another harmonic oscillator parameter $\alpha_{\lambda}$ is determined by $\alpha_{\lambda}=(\frac{3m'}{2m+m'})^{1/4}\alpha_{\rho}$, where $m'$ denotes the heavy quark mass and $m$ denote the light quark mass. The masses of the final baryons and mesons are taken from the PDG~~\cite{Patrignani:2016xqp}. For the global parameter $\delta$, we fix its value the same as our previous works~\cite{Wang:2018fjm,Zhong:2007gp,Zhong:2008kd}, i.e., $\delta=0.557$.

\section{Results and analysis}

Inspired by the newly observed four peaks by the LHCb Collaboration~\cite{Aaij:2020cex}, we carry out a systematic study of the strong decay behaviors of the singly bottom baryons $\Omega_b$ up to $N=2$ shell within the $j$-$j$ coupling scheme in the framework of the chiral quark model. Our results and theoretical predictions are presented as follows.


\subsection{$1P$ states}
The predicted masses of the $\lambda$-mode $1P$-wave baryons $\Omega_b$ are about 6.3 GeV (see Table~\ref{mass-spectrum}), which are in good agreement with the masses of the newly observed four $\Omega_b$ states~\cite{Aaij:2020cex}. In addition, via evaluation the OZI-allowed two-body decays, some people obtained that the $1P$ $\Omega_b$ states have a narrow decay widths~\cite{Santopinto:2018ljf,Agaev:2017ywp}, which are in accordance with the nature of the newly observed $\Omega_b$ states. To pin down the inner structures of the four peaks, it is crucial to study the decay properties of those $1P$ states carefully.

\subsubsection{$J^P=1/2^-$ states}

\begin{figure}[ht]
\centering \epsfxsize=8.5 cm \epsfbox{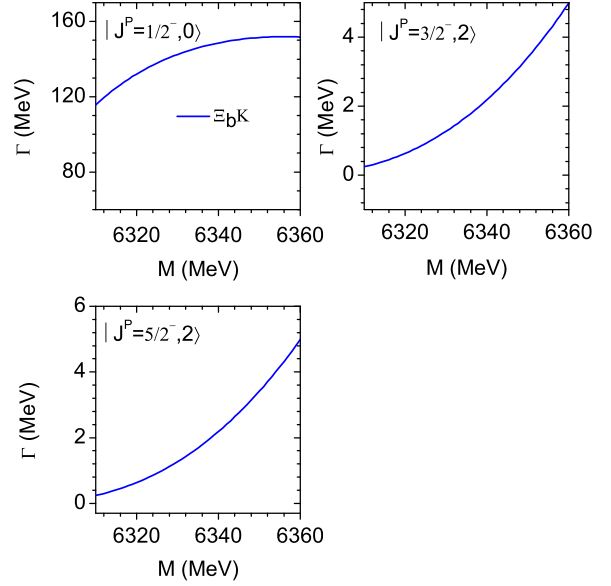} \vspace{-0.6 cm} \caption{Partial decay widths of the $\Xi_b K$ channel for the $J^P = 1/2^-$, $J^P = 3/2^-$ and $J^P = 5/2^-$ $\Omega_b$ states as a function of the mass.}\label{p-mass}
\end{figure}

There are two $J^P=1/2^-$ states $|J^P=\frac{1}{2}^-,0\rangle $ and $|J^P=\frac{1}{2}^-,1\rangle $ within the $j$-$j$ coupling scheme (see Table~\ref{JJ}). They can be expressed as mixed states between $|^2P_{\lambda}\frac{1}{2}^-\rangle$ and $|^4P_{\lambda}\frac{1}{2}^-\rangle$:
\begin{eqnarray}
\left|J^P=\frac{1}{2}^-,0\right\rangle=-\sqrt{\frac{1}{3}}\left|^2P_{\lambda}\frac{1}{2}^-\right\rangle+\sqrt{\frac{2}{3}}\left|^4P_{\lambda}\frac{1}{2}^-\right\rangle,\\
\left|J^P=\frac{1}{2}^-,1\right\rangle=\sqrt{\frac{2}{3}}\left|^2P_{\lambda}\frac{1}{2}^-\right\rangle+\sqrt{\frac{1}{3}}\left|^4P_{\lambda}\frac{1}{2}^-\right\rangle.~~~
\end{eqnarray}
Their masses are predicted to be above the threshold of $\Xi_b K$. Considering the uncertainties of the predicted mass, we plot the variation of the decay width of the $\Xi_b K$ decay mode as a function of the mass of the state $|J^P=1/2^-,0\rangle$ in Fig.~\ref{p-mass}. The decay width increases with the mass increasing in the range of (6290-6350) MeV, and varies in the region of   $\Gamma\sim(33-151)$ MeV. The decay properties of this state is inconsistent with the natures of the four $\Omega_b$ states, thus, the four $\Omega_b$ states as the state $|J^P=1/2^-,0\rangle$ should be excluded. Since the predicted width of $|J^P=1/2^-,0\rangle$ is not broad, this state might be observed in the $\Xi_b K$ channel when enough data are accumulated in experiments.

\begin{table}[htp]
\begin{center}
\caption{\label{owmigeP}  Partial decay widths (MeV) of the $1P$ states within the $j$-$j$ coupling scheme in
the $\Omega_b$ family. The value inside of the braces denotes the reference mass (MeV) of the corresponding state.}
\begin{tabular}{p{2.5cm}|p{2.5cm}p{2.5cm}ccccccccccccccccccccccccccc}\hline\hline
State            &$~~~|J^P,j\rangle$   &~~$\Gamma[\Xi_bK]$ (MeV) ~~     \\ \hline
$\Omega_b(6316)$      &$|J^P=\frac{1}{2}^-,1\rangle$   &~~~~~~$\cdot\cdot\cdot$             \\
$\Omega_b(6316)$      &$|J^P=\frac{1}{2}^-,0\rangle$   &~~~~~~126                        \\
$\Omega_b(6340)$      &$|J^P=\frac{3}{2}^-,2\rangle$   &~~~~~~2.2                      \\
$\Omega_b(6330)$     &$|J^P=\frac{3}{2}^-,1\rangle$   &~~~~~~$\cdot\cdot\cdot$      \\
$\Omega_b(6350)$      &$|J^P=\frac{5}{2}^-,2\rangle$   &~~~~~~3.4                  \\
\hline
\end{tabular}
\end{center}
\end{table}

\begin{figure}[ht]
\centering \epsfxsize=8.5 cm \epsfbox{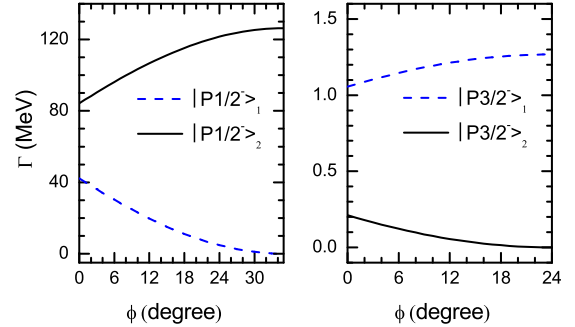}  \caption{Partial decay widths for the $J^P = 1/2^-$ and $J^P = 3/2^-$ $\Omega_b$ states as a function of the mixing angle.}\label{p-degree}
\end{figure}

For the other $J^P=1/2^-$ state $|J^P=1/2^-,1\rangle$, the decay mode $\Xi_b K$ is forbidden in the heavy quark symmetry limit (see Table~\ref{owmigeP}), thus, this state should be very narrow state. We notice that the widths of two peaks, $\Omega_b(6316)$ and $\Omega_b(6330)$, observed in the $\Xi^0_b K^-$ mass spectra~\cite{Aaij:2020cex} are $0.00^{+0.7}_{-0.0}$ MeV and $0.00^{+0.4}_{-0.0}$ MeV, respectively, which are good candidates of this state.

Considering the heavy-quark symmetry is only an approximation, the mixing angle between two states within the $L$-$S$ coupling scheme should be slightly different from that determined in the $j$-$j$ coupling scheme. Namely, the mixing angle $\phi$ for the physical states as mixed states between $|^2P_{\lambda}J\rangle$ and $|^4P_{\lambda}J\rangle$ within the $L$-$S$ coupling scheme, i.e.,
\begin{equation}\label{mixd}
\left(\begin{array}{c}| P~{J^P}\rangle_1\cr |  P~{J^P}\rangle_2
\end{array}\right)=\left(\begin{array}{cc} \cos\phi & \sin\phi \cr -\sin\phi &\cos\phi
\end{array}\right)
\left(\begin{array}{c} |^2P_{\lambda}J^P
\rangle \cr |^4P_{\lambda}J^P\rangle
\end{array}\right),
\end{equation}
may rang from the value of the $L$-$S$ coupling scheme ($\phi=0^\circ$) to that of the $j$-$j$ coupling scheme.

For the $P$-wave states with $J^P=1/2^-$, the mixing angle obtained in the heavy quark limit within the $j$-$j$ coupling scheme is $\phi\simeq35^\circ$, namely $\cos\phi=\sqrt{2/3}$ and $\sin\phi=\sqrt{1/3}$. Thus according to the mixing scheme defined in Eq.~(\ref{mixd}), in Fig.~\ref{p-degree} we plot the strong decay widths of the $J^P=1/2^-$ states as a function of the mixing angle $\phi$ in the region of $(0^\circ,35^\circ)$ by fixing the mass of the resonance at $M=6316$ MeV. It is found that the decay widths are strongly dependent on the mixing angle. The decay width of $|P \frac{1}{2}^-\rangle_2$ is about $\Gamma\sim(83-126)$ MeV with the mixing angle varied in the region what we considered. For the the other mixed state $|P \frac{1}{2}^-\rangle_1$, the decay width is close to zero with the mixing angle tending to $\phi\simeq35^\circ$ which is determined in the $j$-$j$ coupling scheme. This may be evidence that the heavy quark symmetry is a good approximation for the physical states of the singly bottom baryons, thus, the properties of the newly observed $\Omega_b$ states, $\Omega_b(6316)$ and $\Omega_b(6330)$, can be naturally understood.

\subsubsection{$J^P=3/2^-$ states}

Within the $j$-$j$ coupling scheme, there are two $J^P=3/2^-$ $P$-wave states $|J^P=\frac{3}{2}^-,1\rangle$ and $|J^P=\frac{3}{2}^-,2\rangle$, which can be expressed as mixed states in the $L$-$S$ coupling scheme by Eq.~(\ref{Rela}):
\begin{eqnarray}
\left|J^P=\frac{3}{2}^-,1\right\rangle=-\sqrt{\frac{1}{6}}\left|^2P_{\lambda}\frac{3}{2}^-\right\rangle+\sqrt{\frac{5}{6}}\left|^4P_{\lambda}\frac{3}{2}^-\right\rangle,\\
\left|J^P=\frac{3}{2}^-,2\right\rangle=\sqrt{\frac{5}{6}}\left|^2P_{\lambda}\frac{3}{2}^-\right\rangle+\sqrt{\frac{1}{6}}\left|^4P_{\lambda}\frac{3}{2}^-\right\rangle.~~~
\end{eqnarray}
For the two $J^P=3/2^-$ states, the $\Xi_b K$ decay mode is the only OZI-allowed two-body strong decay channel. The decay mode $\Xi_b K$ is forbidden for the state $|J^P=\frac{3}{2}^-,1\rangle$ in the heavy-quark symmetry limit, thus, this state should be very narrow state. Combining the natures of the newly observed states $\Omega_b(6316)$ and $\Omega_b(6330)$, we obtain that these two $\Omega_b$ resonances are good candidates of the $|J^P=\frac{3}{2}^-,1\rangle$ state as well.

For the other $J^P=3/2^-$ state $|J^P=\frac{3}{2}^-,2\rangle$, we also plot the decay width as a function of the mass in the range of $M=(6290-6350)$ MeV in Fig.~\ref{p-mass}. It is found that the decay width is very narrow as well.
Fixing the mass of $|J^P=\frac{3}{2}^-,2\rangle$ at $M=6340$ MeV, we obtain
\begin{eqnarray}
\Gamma[|J^P=\frac{3}{2}^-,2\rangle\rightarrow\Xi_bK]\simeq2.2~\text{MeV}.
\end{eqnarray}
This theoretical width is well comparable with the observations of the two resonances $\Omega_b(6340)$ and $\Omega_b(6350)$~\cite{Aaij:2020cex}, thus the two resonances are good candidates of the $J^P=3/2^-$ state $|J^P=\frac{3}{2}^-,2\rangle$.

The mixing angle for the $J^P=3/2^-$ physical states may be slightly different from the value $\phi=24^\circ$ determined by the $j$-$j$ coupling scheme in the heavy-quark symmetry limit. Similarly we fix the masses at $M=6330$ MeV and plot the variation of the strong decay widths of the $J^P=3/2^-$ states as a function of the mixing angle in Figure~\ref{p-degree}. From the figure, it is found that the decay widths for the two $J^P=3/2^-$ states are slightly depend on the mixing angle and less than $\Gamma<1.3$ MeV in the range of ($0^\circ$, $24^\circ$).

\subsubsection{$J^P=5/2^-$ states}

There is only one $P$-wave state with $J^P=5/2^-$. There is no differences between the $j$-$j$ coupling scheme and $L$-$S$ coupling scheme, namely
\begin{eqnarray}
\left|J^P=\frac{5}{2}^-,2\right\rangle =\left|^4P_{\lambda}\frac{5}{2}^-\right\rangle.
\end{eqnarray}
According to the Table~\ref{mass-spectrum}, the theoretical masses of the $J^P=5/2^-$ states are about 6.3 GeV. Thus, we plot the decay width of this state as a function of the mass in Fig.~\ref{p-mass} in the region of (6290-6350) MeV. It is found that this state is most likely narrow ($\Gamma<3.5$ MeV) within the mass range what we considered, and comparable with the nature of the newly observed $\Omega_b$ states, $\Omega_b(6340)$ and $\Omega_b(6350)$. Fixing the mass of $|J^P=\frac{5}{2}^-,2\rangle$ at $M=6350$ MeV, we obtain
\begin{eqnarray}
\Gamma[|J^P=\frac{5}{2}^-,2\rangle\rightarrow\Xi_bK]\simeq3.4~\text{MeV}.
\end{eqnarray}

So far, we have calculated the strong decay properties of the five $\lambda$-mode $P$-wave $\Omega_b$ states within the $j$-$j$ coupling scheme. Our results show that excepting the state $|1/2^-,0\rangle$ being a middle width state with a width of $\Gamma\sim100$ MeV, the other four states are most likely to be narrow states, which are good candidates for the newly observed four $\Omega_b$ states. The newly observed resonances $\Omega_b(6316)$ and $\Omega_b(6330)$ can be interpreted as the light spin $j=1$ states, while their spin-parity $J^P=1/2^-$ and $J^P=3/2^-$ cannot be distinguished. The other two states $\Omega_b(6340)$ and $\Omega_b(6350)$ may be the light spin $j=2$ states, but the spin-parity $J^P=3/2^-$ and $J^P=5/2^-$ cannot be distinguished.

\subsection{$2S$ states}

According to the predicted masses of the $2S$ states in the $\Omega_b$ family (see Table~\ref{mass-spectrum}), the possibility as the candidates of the newly observed resonances cannot be excluded completely. It is necessary to study the strong decay properties of the $2S$ states.

There are two $2S$ states with spin-parity $J^P=1/2^+$ and $j^P=3/2^+$, respectively. For these two 2S-wave states, there is no differences between the $j$-$j$ coupling scheme and $L$-$S$ coupling scheme. Adopting the predicted masses from the Ref.~\cite{Ebert:2011kk}, we calculate the partial decay widths of the two $2S$ states. Our results have been listed in Table~\ref{omega2S}. From the table, we get that the two $2S$ states may be narrow state with a comparable total decay width of $\Gamma\sim 1$ MeV.

\begin{table}[htb]
\begin{center}
\caption{ \label{omega2S} Partial decay widths (MeV) for the $2S$-wave $\Omega_b$ states, whose masses (MeV) are taken the quark model predictions of Ref.~\cite{Ebert:2011kk}.}
\begin{tabular}{c|ccccc}
\hline\hline
 \multirow{2}{*}{Decay mode}        &\multicolumn{1}{c}{$\underline{~~~~~~|J=\frac{1}{2}^+,1\rangle(2S)~~~~~~}$} &\multicolumn{1}{c}{$\underline{~~~~~~|J=\frac{3}{2}^+,1\rangle(2S)~~~~~~}$} \\
             &~~ $\Omega_b(6450)$            &~~ $\Omega_b(6461)$         \\ \hline
 $\Xi_b K$                      &0.51                       &0.36     \\
 $\Xi'_b K$                       &0.32        &0.14     \\
 $\Xi_b^*K$                     &0.04        &0.10      \\
Sum                                  &0.87                        &0.60      \\
\hline
\hline
\end{tabular}
\end{center}
\end{table}

\begin{figure}[ht]
\centering \epsfxsize=4.8 cm \epsfbox{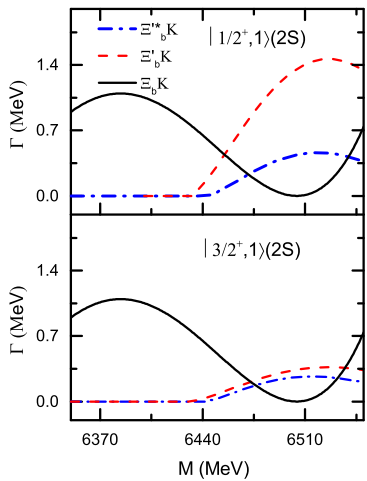}  \caption{Partial decay widths of the $2S$ states with $J^P = 1/2^+$ and $J^P = 3/2^+$, respectively, as a function of the masses.}\label{2S-mass}
\end{figure}

Considering the uncertainty of the predicted masses, we plot the strong decay properties of the two states as functions of the masses in the range of (6.35,6.55) MeV, as shown in Fig.~\ref{2S-mass}. The total decay widths of the $2S$ states are about $\Gamma<2$ MeV within the masses we considered.

The decay properties of the $2S$ states were calculated in Ref.~\cite{Liang:2020hbo} within quark pair creation model as well. The authors predicted that the total decay widths of the two $2S$ states were about $\Gamma\sim50$ MeV, which were incomparable with our predictions.

It should be pointed out that although some people obtain the theoretical masses of the $2S$ states in the $\Omega_b$ family are about $6.3-6.4$ GeV~\cite{Yang:2017qan,Garcilazo:2007eh} and the possibility as the new observed $\Omega_b$ states cannot be excluded absolutely, however, the mass splitting between the $1P$ and $2S$ is about $100-150$ MeV. We notice that the splitting between the four newly observed resonances ranges from several to 10 MeV. If the four resonances are conventional baryons and two of them are interpreted as the $2S$ states, while how to understand the other two resonances will be a great challenge.

\begin{table*}[!]
\begin{center}
\caption{\label{omigeD} Partial widths (MeV) of strong decays for the $\lambda$-mode $D$-wave  $\Omega_b$
baryons, whose masses (MeV) are taken the quark model predictions of Ref.~\cite{Ebert:2011kk}. }
\scalebox{1.0}{
\begin{tabular}{cc|ccccccccccccccc}\hline\hline
\multirow{2}{*}{Decay mode}
 &\multirow{2}{*}{$M_f$}
&\multicolumn{2}{c}{$\underline{|J=\frac{1}{2}^+,1\rangle(6540)}$}
&\multicolumn{2}{c}{$\underline{|J=\frac{3}{2}^+,1\rangle(6530)}$}
&\multicolumn{2}{c}{$\underline{|J=\frac{3}{2}^+,2\rangle(6549)}$}
&\multicolumn{2}{c}{$\underline{|J=\frac{5}{2}^+,2\rangle(6520)}$}
&\multicolumn{2}{c}{$\underline{|J=\frac{5}{2}^+,3\rangle(6529)}$}
&\multicolumn{2}{c}{$\underline{|J=\frac{7}{2}^+,3\rangle(6517)}$}       \\
& &$\Gamma_i$   &$\mathcal{B}_i$ (\%)   &$\Gamma_i$   &$\mathcal{B}_i$ (\%)    &$\Gamma_i$   &$\mathcal{B}_i$ (\%)   &$\Gamma$   &$\mathcal{B}_i$ (\%)   &$\Gamma_i$   &$\mathcal{B}_i$ (\%)   &$\Gamma_i$   &$\mathcal{B}_i$ (\%)              \\
\hline
 $\Xi_bK$               &  5794    &22.1       &77.1       &22.4     &81.5        &$\cdot\cdot\cdot$        &$\cdot\cdot\cdot$      &$\cdot\cdot\cdot$       &$\cdot\cdot\cdot$      &9.06      &91.5     &7.83         &91.05        \\
 $\Xi_b^{'}K$           &  5935    &4.69       &16.4       &1.05     &3.8      &9.49      &85.3     &0.15    &2.6      &0.24    &2.4      &0.09         &1.04         \\
 $\Xi_b^{*}K$           &  5955    &1.86       &6.5       &4.04     &14.7       &1.63     &14.7     &5.61   &97.4      &0.60    &6     &0.68         &7.91          \\
 total                  &          &28.65      &          &27.49     &      &11.12       &          &5.76  &       &9.90      &     &8.6 &\\
\hline\hline
\end{tabular}}
\end{center}
\end{table*}

\subsection{$1D$ states}

The $1D$-wave $\Omega_b$ states are most likely to be observed in forthcoming experiments as well. In Ref.~\cite{Yao:2018jmc}
we have studied the strong decays of the $1D$-wave $\Omega_b$ states in the $L$-$S$ coupling scheme. Since the heavy-quark symmetry
indicates the the singly-heavy baryons may more favor the $j$-$j$ coupling scheme, in present work we further give
our predictions in the $j$-$j$ coupling scheme.

For the $\lambda$-mode $1D$ states, there are six states within the $j$-$j$ coupling scheme and the corresponding relations to the states in the $L$-$S$ coupling scheme are
\begin{eqnarray}
\left|J^P=\frac{1}{2}^+,1\right\rangle=&\left|^4D_{\lambda\lambda}\frac{1}{2}^+\right\rangle,~~~~~~~~~~~~~~~~~~~~~~~~~~~~~~~~~~~~~~~\\
\left|J^P=\frac{3}{2}^+,1\right\rangle=&-\sqrt{\frac{1}{2}}\left|^2D_{\lambda\lambda}\frac{3}{2}^+ \right\rangle + \sqrt{\frac{1}{2}}\left|^4D_{\lambda\lambda}\frac{3}{2}^+ \right\rangle,~~~~~~~\\
\left|J^P=\frac{3}{2}^+,2\right\rangle=&\sqrt{\frac{1}{2}}\left|^2D_{\lambda\lambda}\frac{3}{2}^+ \right\rangle + \sqrt{\frac{1}{2}}\left|^4D_{\lambda\lambda}\frac{3}{2}^+ \right\rangle,~~~~~~~~~\\
\left|J^P=\frac{5}{2}^+,2\right\rangle=&-\frac{\sqrt{2}}{3}\left|^2D_{\lambda\lambda}\frac{5}{2}^+ \right\rangle + \frac{\sqrt{7}}{3}\left|^4D_{\lambda\lambda}\frac{5}{2}^+ \right\rangle,~~~~~~~\\
\left|J^P=\frac{5}{2}^+,3\right\rangle=&\frac{\sqrt{7}}{3}\left|^2D_{\lambda\lambda}\frac{5}{2}^+ \right\rangle + \frac{\sqrt{2}}{3}\left|^4D_{\lambda\lambda}\frac{5}{2}^+ \right\rangle,~~~~~~~~\\
\left|J^P=\frac{7}{2}^+,3\right\rangle=&\left|^4D_{\lambda\lambda}\frac{7}{2}^+ \right\rangle.~~~~~~~~~~~~~~~~~~~~~~~~~~~~~~~~~~~~~~~
\end{eqnarray}
For the $D$-wave $\Omega_b$ states, within the masses adopted from Ref.~~\cite{Ebert:2011kk} we calculate the strong decays of the $1D$ states and collect in Table~\ref{omigeD}.

From Table~\ref{omigeD}, it is seen that the total decay widths of $|J=1/2^+,1\rangle$ and $|J=3/2^+,1\rangle$ are about $\Gamma\sim28$ MeV, and both mainly decay into $\Xi_bK$. The strong decays of the states  $|J=5/2^+,3\rangle$ and $|J=7/2^+,3\rangle$ are governed by the $\Xi_bK$ channel as well, and the total decay widths of both states are about $\Gamma\sim10$ MeV.
However, the decay mode  $\Xi_bK$ is forbidden for the states $|J=3/2^+,2\rangle$ and $|J=5/2^+,2\rangle$. The two states might be narrow state with the total decay widths of $\Gamma\sim11$ MeV and $\Gamma\sim6$ MeV, respectively. Meanwhile, we notice that the strong decays of the state $|J=3/2^+,2\rangle$ is dominated by the $\Xi'_bK$ channel, while that of the state $|J=5/2^+,2\rangle$ is the $\Xi^*_bK$ channel.

Comparing with the predictions in Ref.~\cite{Liang:2020hbo}, the main decay channel for each $1D$ states is consistent with each other. However the total decay widths are difference, especially for the states $|J=1/2^+,1\rangle$ and $|J=3/2^+,1\rangle$. The predicted total decay widths for the two states in Ref.~\cite{Liang:2020hbo} are about three times larger than ours.

\section{Summary}

In this paper, we carry out a systematic study of the OZI allowed two-body strong decays of the low-lying $\lambda$-mode $\Omega_b$ resonances up to the $N=2$ shell in the framework of a chiral quark model within the $j$-$j$ coupling scheme. For the newly observed four $\Omega_b$ states, we give a possible interpretation in theory. Meanwhile, we give the predictions for the decay properties of the $2S$- and $1D$- wave states, and hope to provide helpful information for searching these missing $\Omega_b$ states in the future.

Our theoretical results show that the newly observed states $\Omega_b(6316)$ and $\Omega_b(6330)$ are most likely to be explained as the $1P$ wave states with the light spin $j=1$, while the spin-parity $J^P=1/2^-$ and $J^P=3/2^-$ cannot be distinguished. The other two newly observed states $\Omega_b(6340)$ and $\Omega_b(6350)$ may correspond to the $1P$-wave states with the light spin $j=2$ excepting the spin-parity $J^P=3/2^-$ and $J^p=5/2^-$ cannot being distinguished. In addition, we notice that the state $|1/2^-,0\rangle(1P)$ might be a moderate width state with a total decay width of $\Gamma\sim100$ MeV, and the $\Xi_bK$ channel mostly saturates its total decay width.

For the two $2S$ states with the spin-parity $J^P=1/2^+$ and $J^P=3/2^+$, they might be narrow states with a total width of $\Gamma<2$ MeV. Considering the large mass splitting between $1P$- and $2S$-wave states together with the small mass splitting along the observed four peaks, it is unnatural to explain two of the four peaks as $2S$-wave states within the conventional baryon picture.

For the $1D$ states, the total decay widths are not broad and vary from several to dozens of MeV. In addition, we notice that the strong decays of the states $|J=1/2^+,1\rangle$, $|J=3/2^+,1\rangle$, $|J=5/2^+,3\rangle$, and $|J=7/2^+,3\rangle$ are governed by the $\Xi_bK$ channel, while this channel is forbidden for the states $|J=3/2^+,2\rangle$ and $|J=5/2^+,2\rangle$. The state $|J=3/2^+,2\rangle$ dominantly decays to the $\Xi'_bK$ channel, and the state $|J=5/2^+,2\rangle$ are governed by the $\Xi^*_bK$ channel.

\section*{Acknowledgements }
This work is supported by the National Natural
Science Foundation of China under Grants No. 11775078, No. 11947048 and  No. U1832173.


\end{document}